# A Continuous Liveness Detection System for Text-independent Speaker Verification

Linghan Zhang[1], Sheng Tan[1], Yingying Chen[2], and Jie Yang[1]

[1]Florida State University; [2]Rutgers University

**Abstract**—Voice authentication is drawing increasing attention and becomes an attractive alternative to passwords for mobile authentication. Recent advances in mobile technology further accelerate the adoption of voice biometrics in an array of diverse mobile applications. However, recent studies show that voice authentication is vulnerable to replay attacks, where an adversary can spoof a voice authentication system using a pre-recorded voice sample collected from the victim. In this paper, we propose VoiceLive, a liveness detection system for both text-dependent and text-independent voice authentication on smartphones. VoiceLive detects a live user by leveraging the user's unique vocal system and the stereo recording of smartphones. In particular, utilizing the built-in gyroscope, loudspeaker and microphone, VoiceLive first measures the smartphone's distance and angle from the user, then it captures the position specific time-difference-of-arrival (TDoA) changes in a sequence of phoneme sounds to the two microphones of the phone, and uses such unique TDoA dynamic which doesn't exist under replay attacks for liveness detection. VoiceLive is practical as it doesn't require additional hardware but two-channel stereo recording that is supported by virtually all smartphones. Our experimental evaluation with 12 participants and different types of phones shows that VoiceLive achieves over 99% detection accuracy at around 1% Equal Error Rate (EER) on the text-dependent system and around 99% accuracy and 2% EER on the text-independent one. Results also show that VoiceLive is robust to different phone positions, i.e. the user are free to hold the smartphone with distinct distances and angles.

## 1 Introduction

As a primary way of communication, our voice is a particularly attractive biometric for identifying users. It reflects individual differences in both behavioral and physiological characteristics, such as the inflection and the shape of the vocal tract [27]. Such distinctive behavioral and physiological traits could be captured by voice authentication systems for differentiating each individual [21]. Voice authentication leveraging built-in microphones on mobile devices is particularly convenient and low-cost, comparing to the passwords authentication that is difficult to use while on-the-go and requires memorization. Recent advances in mobile technology further accelerate the adoption of voice biometrics in an array of diverse mobile applications [46].

Indeed, voice authentication has been introduced recently to mobile devices and apps to provide secure access and logins. For example, Google has integrated it into Android operating systems (OSs) to allow users to unlock mobile devices [2], and Tencent has updated its WeChat mobile app to support voice biometric logins [9]. Another appealing use case of voice authentication is to support mobile financial services. For instance, SayPay provides voice biometric solution for online payment, e-commerce, and online banking [4]. And an increasing number of financial institutions, HSBC, Citi, and Barclays for example, are deploying voice authentication for their telephone and online banking systems [3]. This market is expected to continue growing at an annual growth rate of 17.2% over 2020 to 2025, and will reach $26.8 billion by 2025 [5]. Voice authentication thus becomes an attractive alternative to passwords in mobile authentication and is increasingly popular.

Voice authentication however has been shown to be vulnerable to replay attacks in recent studies [20]. An adversary can spoof a voice authentication system by using a pre-recorded voice sample collected from the victim. The voice sample can be any recording captured inconspicuously. Or, an adversary can obtain voice samples from the victim's publicly exposed speeches. The attacker could even concatenate voice samples from a number of segments in order to match the victim's passphrase. Such attacks are most accessible to the adversary due to the proliferation of mobile devices, such as smartphones and digital recorders. They are also highly effective in spoofing authentication systems, as evidenced by recently work [37], [39]. Replay attacks therefore present significant threats to voice authentication and are drawing increasing attention. For example, Google advises users on the vulnerability of their voice logins by displaying a popup message "... a recording of your voice could unlock your device." [1]

Prior work in defending against replay attacks is to uti-

- *L. Zhang and J. Yang are with the Department of Computer Science, Florida State University, Tallahassee, FL, 32306.*
- *Y. Chen is with the Department of Electrical and Computer Engineering, Rutgers University, New Brunswick, NJ, 08901.*
- *S. Tan is with the Department of Computer Science, Trinity University, San Antonio, TX, 78212.*

lize liveness detection to distinguish between a passphrase spoken by a live user and a replayed one pre-recorded by the adversary. For example, Shang *et al.* propose to compare an input voice sample with stored instances of past accesses to detect the voice samples have been seen before by the authentication system [36]. This method, however, cannot work if the attacker records the voice samples during a non-authentication time point. Villalba *et al.* and Wang *et al.* suggest that the additional channel noises introduced by the recording and loudspeaker can be used for attack detection [37], [39]. These approaches however have limited effectiveness in practice. Moreover, recent liveness detection solutions measure unique features of a live user while he speaks [18], [30], [35], [38], [44]. These methods, however, are limited by the distance between the user and the device. Indeed, they all require the user to carry or wear addition devices, or to stay close to the device. Furthermore, these work are text-dependent, which only function on enrolled passphrases. Although CaField achieves text-interdependency and reduces such position constraints, it necessitates consistent fixed location of the smartphone [42].

In this paper, we introduce and evaluate a phoneme sound localization based liveness detection system on smartphones. Our system distinguishes a passphrase spoken by a live user from a replayed one by leveraging (i) the human speech production system and (ii) advanced smartphone audio hardware. First, in human speech production, a phoneme is the smallest distinctive unit sound of a language. Each phoneme sound can be viewed as air waves produced by the lungs, and then modulated by the movements of vocal cords and vocal tract including throat, mouth, nose, tongue, teeth, and lips. Each phoneme sound thus experiences unique combination of place and manner of articulation. Consequently, different phoneme sounds could be located at different physical positions in the human vocal tract system with an acoustic localization method. Second, smartphone hardware is now supporting advanced audio capabilities. Virtually all smartphones are equipped with two microphones for stereo recording (one on the top and the other one at the bottom), and are capable of recording at standard 48kHz and 192kHz sampling rates. For example, with the latest Android OSs, Samsung Galaxy S5 and Note3 are capable of stereo recording at 192kHz, which yields 5.21 microseconds' time resolution or millimeter-level ranging resolution[1]. We thus can leverage such stereo recording or dual microphones on smartphones to pinpoint the sound origin of each phoneme within human vocal system for liveness detection.

Ideally, locating a phoneme sound requires at least three microphones with three individual audio channels. Although current two-channel stereo recording cannot uniquely locate the phoneme sound origin, it can capture the time-difference-of-arrival (TDoA) of each phoneme sound to the two microphones of the phone. Indeed, the differences in TDoA between most phoneme sounds are distinctive and measurable with millimeter-level ranging resolution. Very importantly, each passphrase (usually 5 to 7 words [8], [34]) consists of a sequence of different phoneme sounds that will produce a series of TDoA measurements with various values. We refer to the changes in TDoA values as TDoA dynamic, which is determined by the specific passphrase, the placement of the phone, and a user's unique vocal system. Such TDoA dynamic, which doesn't exist under replay attacks, is then utilized for liveness detection.

In particular, for the text-dependent liveness detection system, when a user first enrolled in the system, the TDoA dynamic of the user-chosen or system prompted passphrase is first captured by the smartphone stereo recording, and then stored in the system. During online authentication phase, the extracted TDoA dynamic of an input utterance will be compared to the one stored in the system. For the text-independent liveness detetion system, instead of enrolling a new user with specific passphrases, VoiceLive collects all the phonetic sounds of a language and extracts the corresponding TDoAs as templates. When authenticating the user, VoiceLive recognizes phonetic sounds and searches for the corresponding phoneme TDoA templates, which, are then combined to form the speech TDoA template the user speaks. This template is them adopted for similarity comparison against the TDoA dynamic extracted from the speech. A live user is detected, if that produce a similarity score higher than a pre-defined threshold. By relaxing the problem from locating each phoneme sound to measuring the TDoA dynamic for a sequence of phonemes, we enable liveness detection on a single phone without any additional hardware. Considering TDoA measurements vary with different positions of the smartphone, the user must hold the phone close to her/his mouth with the same pose in both enrollment and authentication processes to ensure similar TDoA dynamics. Fortunately, VoiceLive releases such constraint by building a geometric model that calculates the TDoA changes when the user holds the phone at different locations with inconsistent angles. The contributions of our work are summarized as follows:

- We show that the origin of each phoneme can be uniquely located within the human vocal tract system by using a microphone array. It lays the foundation of our phoneme localization based liveness detection system.
- We develop VoiceLive, a practical liveness detection system that extracts the TDoA dynamic of the passphrase for live user detection. VoiceLive takes advantages of the user's unique vocal system and high quality stereo recording of smartphones. VoiceLive could be applied on both text-dependent and text-independent voice authentication systems.
- We conduct extensive experiments with 12 participants and three different types of phones under various experimental settings. Experimental results show that as a text-dependent liveness detection system, VoiceLive achieves over 99% detection accuracy at around 1% EER. For the text-independent experiments, VoiceLive still maintains around 99% accuracy and 2% EER. Results also show that VoiceLive is robust to different phone placements and is compatible to different sampling rates, phone models, and different locations and angles of the smartphones.

The remainder of this paper expands on above contribu-

1. Assuming the speed of sound is 340m/s, each digital sample represents a distance of 1.77mm.

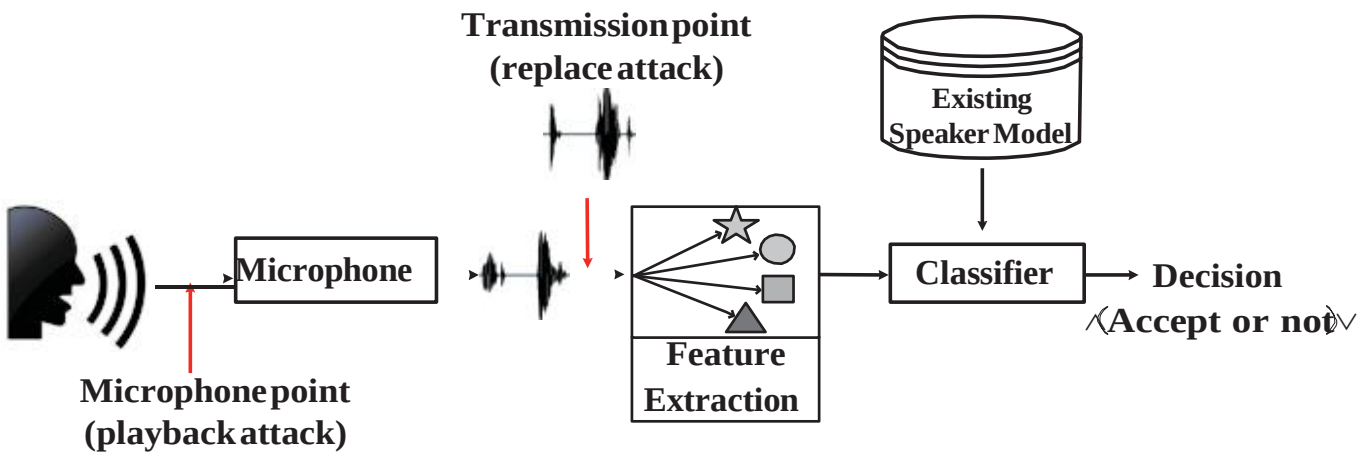

**Fig. 1: A typical voice authentication system with two possible places of replay attacks.**

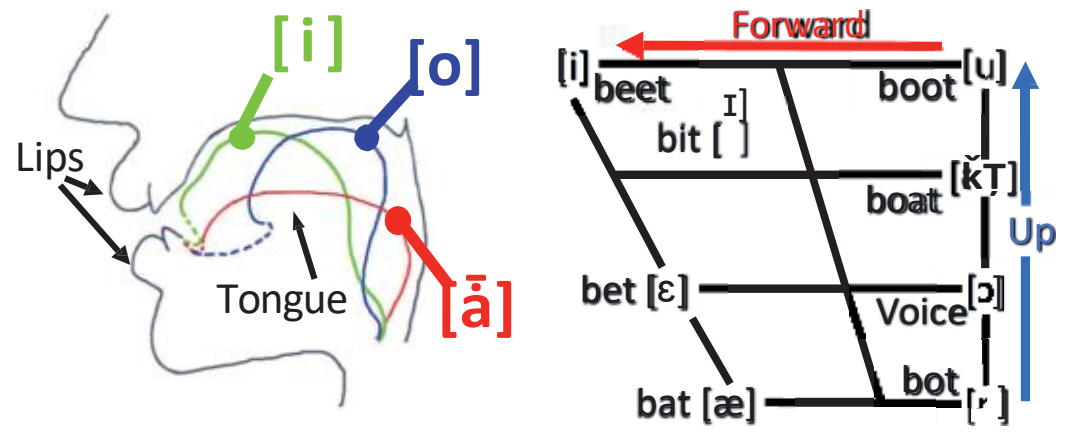

(a) Example (b) Vowel chart

**Fig. 2: Tongue positions of English vowels within the oral cavity, and the vowel chart.**

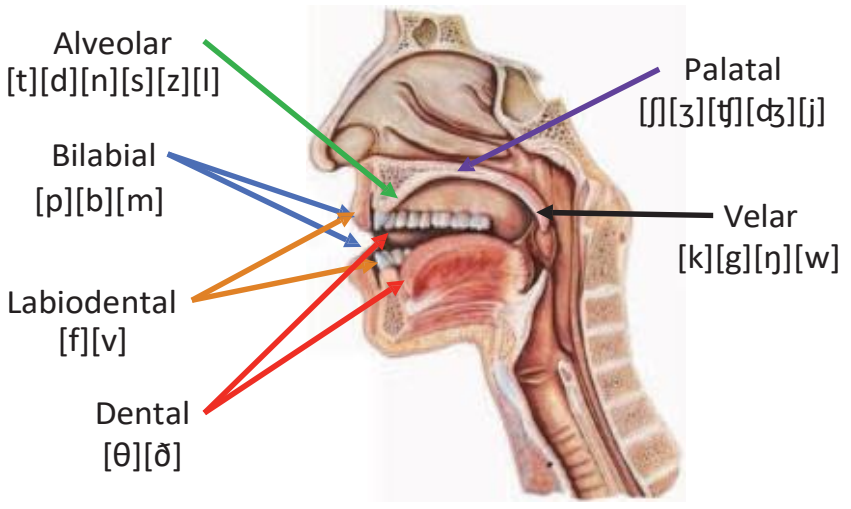

**Fig. 3: Place of articulation and corresponding consonants.**

| Manner \ Place | Bilabial | Labiodental | Dental | Alveolar | Palatal | Velar |
|---|---|---|---|---|---|---|
| Nasal | [m] | | | [n] | | [ŋ] |
| Stop | *[p]* **[b]** | | | *[t]* **[d]** | *[ř]* **[ʒ]** | *[k]* **[g]** |
| Fricative | | *[f]* **[v]** | [θ] **[ð]** | *[s]* **[z]** | | |
| Affricate | | | | | *[ɟ]* **[ʤ]** | |
| Approximate | | | | | | [w] |
| Lateral | | | | [l] | | |

**Fig. 4: Consonants chart based on place and manner of articulation.**

tions. We begin with system and attack model, and a brief introduction to phoneme sounds localization.

## 2 Preliminaries

### 2.1 System and Attack Model

There exists two types of voice authentication systems: text-dependent and text-independent. We primarily focus on the text-dependent system as it is currently the most commercially viable method and produces better authentication accuracy with shorter utterances [36]. In a text-dependent system, the text to be spoken by a user is the same one for enrollment and verification. Such text could be either a user-chosen or system prompted one. Figure 1 shows the processes of a typical voice authentication system. We also extend our method to text-independent system and present the results in Section 4.6.

For the attack model, we consider replay attacks, which are the most accessible and effective attacks aiming at spoofing the system by replaying a pre-recorded voice sample of the victim [37]. We consider the replay attacks that take place at two locations, at the microphone point and at the transmission point, as shown in Figure 1. For the sake of simplicity, we refer to the former as a *playback attack* and the latter as a *replace attack*. In a playback attack, an adversary uses a speaker to replay the pre-recorded voice sample in front of the microphones. In a replace attack, an adversary replaces his/her own speech signal as the victim's before or during transmission. This can be done by leveraging the availability of the virtual recorder to bypass the local microphones, or by intercepting and replacing speech signal during transmission.

### 2.2 Human Speech Production and Phonemes

The human speech production system involves three vital physiological components: lungs, vocal cords, and vocal tract [31]. When someone exhales, air is expelled from the *lungs*, and then passes over the *vocal cords*, which dilate or constrict to allow or impede the air flow to produce unvoiced or voiced sound. Such sound is then resonated and reshaped by the *vocal tract* that consists of multiple organs such as throat, mouth, nose, tongue, teeth, and lips. The vocal cords modulation, interaction and movement of these organs can alter sound waves and produce unique human sounds.

A phoneme is the smallest distinctive unit sound of a language [31]. The two major phoneme categories are vowels and consonants. In particular, vowels are the phoneme sounds produced when vocal cords constrict air flow (i.e., voiced sound) but with an open vocal tract. The tongue position is the most important physical feature that distinguishes one vowel from another [31]. As different tongue positions lead to different multipath environments inside the oral cavity, we can locate the sound origins of different vowels at different physical locations inside the human oral cavity. As illustrated in Figure 2 (a), when the tongue moves to lower right corner, vowel [A] can be pronounced, whereas when the tongue moves to upper left corner and backward, vowels [i] and [o] can be produced, respectively. More generally, Figure 2 (b) shows the vowel chart which involves two dimensions of tongue movements: up/down movements (i.e., height) and back/forth movements (i.e., backness). Extending or retracting the tongue forward or backward towards the teeth produces a more front or back vowel sound, whereas lowering or raising the tongue towards lower jaw or towards the roof of mouth produces a more open or close vowel.

Unlike vowels, consonants are produced when vocal cords either constrict or dilate air flow and with significant constriction of the air flow in the oral cavity. The articulation place and manner are two major factors that distinguish one consonant from another [31]. The combined effect of place and manner of articulation and voiced/unvoiced sound lead to different consonant sounds emitted from different locations within the human vocal tract system. In particular, place of articulation is the location where the constrictions or obstructions of air stream occur, and can be categorized into 6 groups: bilabial, labiodental, dental, alveolar, palatal, and velar. Figure 3 shows each group and the corresponding consonants. For example, the consonants [p][b][m][w] can be pronounced when the obstruction of air stream occurs at upper and lower lips. The consonants within each group can be further distinguished by the manner of articulation, which describes the configuration and interaction of the speech organs (e.g., the tongue, lips, and palate). There are 6 types of articulation manners including nasal, stop, fricative, affricate, approximate and lateral. For instance, nasal consonant [m] is produced when the air stream is completely blocked by mouth and only passes through the nose. Figure 4 summarizes the categorization of different consonants based on place of articulation and manner of articulation. The bolded font in the figure shows the voiced sounds (e.g., *[b]* and *[v]*), whereas the rest are unvoiced sounds (e.g., *[p]* and *[f]*).

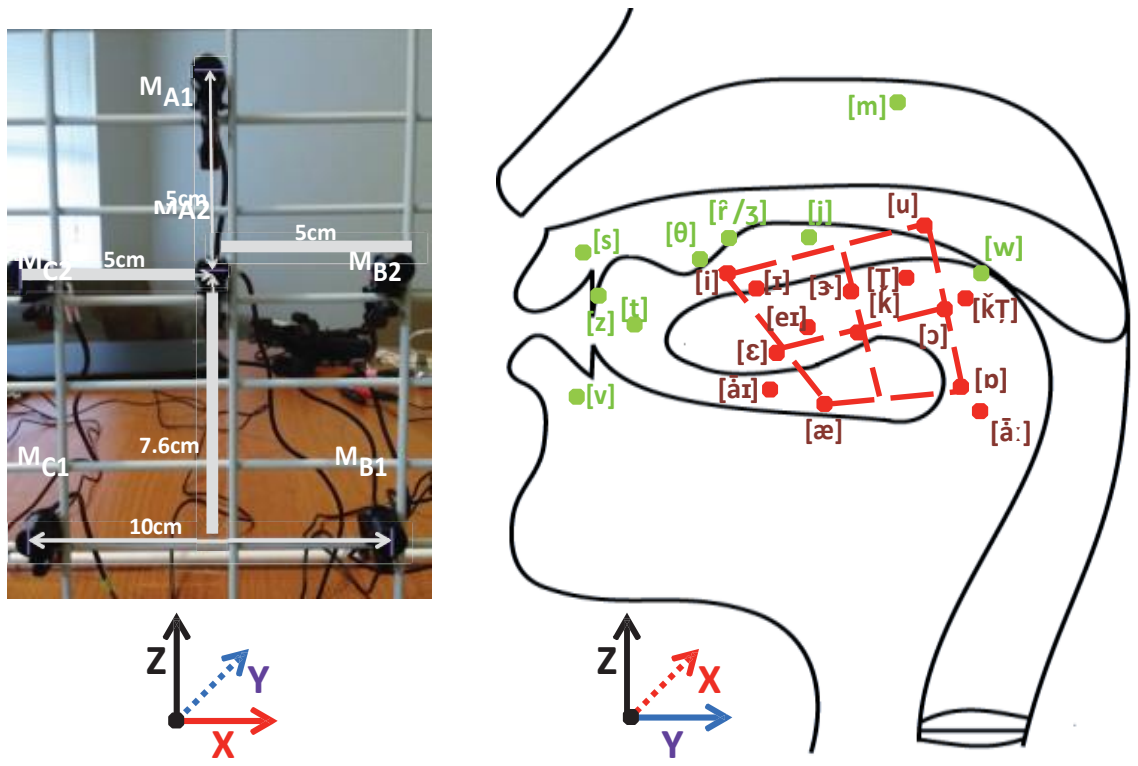

(a) Layout of Microphones (b) Localized phoneme sounds

**Fig. 5: Phonemes localization using microphone array.**

### 2.3 Phoneme Localization using Microphone Array

We next conduct experiments to study how the origin of phoneme sound is located within the human vocal tract system by leveraging a microphone array. We utilize six external microphones organized in three pairs A, B, and C. As shown in Figure 5 (a), the microphones are distributed in the X-Z plane[2] with 5cm and 10cm horizontal distances, and 5cm and 7.6cm vertical distances. Such a distribution could cover the size of a human vocal tract. Each pair is synchronized to measure the TDoA of the sound origin to the two microphones. These pairs produce three independent TDoA values, which could uniquely locate the sound origin in a 3D space. We measure the TDoA in terms of the number of delayed samples to the two microphones. As we use 192kHz for recording, the TDoA ranging resolution is 1.77mm. Before the phoneme localization, we test the localization accuracy by emitting chirp sounds at different fixed locations in front of the microphones. We observe that it produces an averaged localization error within 2mm.

We recruit two participants to pronounce each phoneme sound in front of the microphone array multiple trials. Figure 5 (b) illustrates the localized phoneme sound origins for one participant. It shows the sectional view of human vocal tract on Y-Z plane. The red dots show the localized vowel sound origins, whereas the green dots show these of

2. Note that the sectional view of the human vocal tract in Figure 5 (b) is on the Y-Z plane.

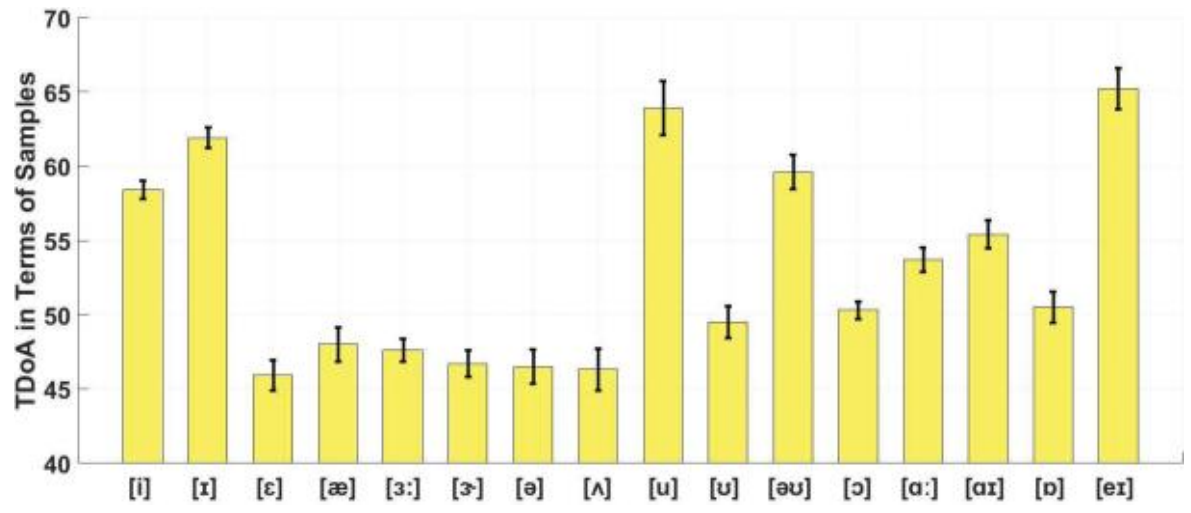

**Fig. 6: An Example of English Vowel Phonetic Sounds.**

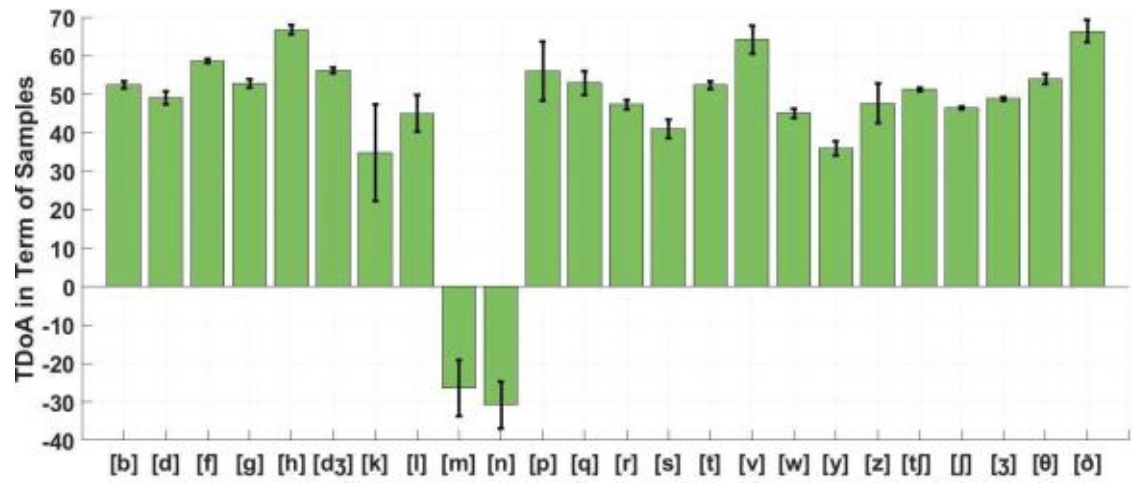

**Fig. 7: An Example of English Consonants Phonetic Sounds.**

consonant ones. We obtain several important observations from Figure 5. First, the located sound origins of vowels match the tongue positions very well. For example, the vowels connected by the dotted lines in Figure 5 (b) have similar relative positions and overall shape as that of the vowel chart in Figure 2 (b). This is because the tongue position is the deterministic factor of vowel production. Second, some of the consonants have the origins close to the place of articulation, while others are significantly affected by the manner of articulation. For instance, [s],[z] and [t] have the localized sound origins close to alveolar, which is the place of articulation of these sounds, whereas [m] is located in the nasal cavity where the airflows out (i.e., manner of articulation). Moreover, we observe the located phoneme origins are mainly distributed within the mouth and nasal cavities with the size of about 4cm by 4cm, and they show little changes in X axis (i.e., lateral direction of mouth). We also find that different participants produce different localized sound origins for the same phoneme due to the individual diversity in the human vocal tract (e.g., shape and size) and the habitual way of pronouncing phonemes.

### 2.4 TDoA Statistics of English Phonetic Sounds

However, a common smartphone equipped with one pair of microphones only has access to one TDoA value, which may not be able to describe the scattered origins of all 44 English phonemes. Moreover, the measured TDoAs are not fixed values since individuals could pronounce the same phoneme with nuanced differences in various contexts. These contexts include distinct language contexts decided by factors like the accent and adjacent phonemes, or external contexts including different recording environments, time, or physical status of the speaker. Therefore, we study the TDoA statistics on the English phonemes.

Figure 6 and Figure 7 show examples of the TDoAs for English vowels and consonants when a user pronounces each of them for 10 times. We collect the data in several sessions that across a month to cover different contexts. During this procedure, the user holds the smartphone(Samsung S8+) vertically at the same location in front of his face at the coordinate (4, -1) on the Y-Z plane, considering the center of the phone as (0, 0). In these Figures, the bars indicate the mean values of the TDoA measurements for different phonemes whereas the error bars show the standard deviations of them. For English vowels, the mean values of these TDoAs range from 46 to 65 samples, and the std (standard deviation) ranges from 1.16 to 2.79 samples; whereas for the consonants, the mean value ranges from 34.8 to 66.8 samples, except for the nasal sounds [m] and [n], whose TDoA measurements are around -30 samples, and the std value ranges from 0.7 to 1.2 samples, except for few voiceless sounds like [k], whose std is as high as 25.2 samples.

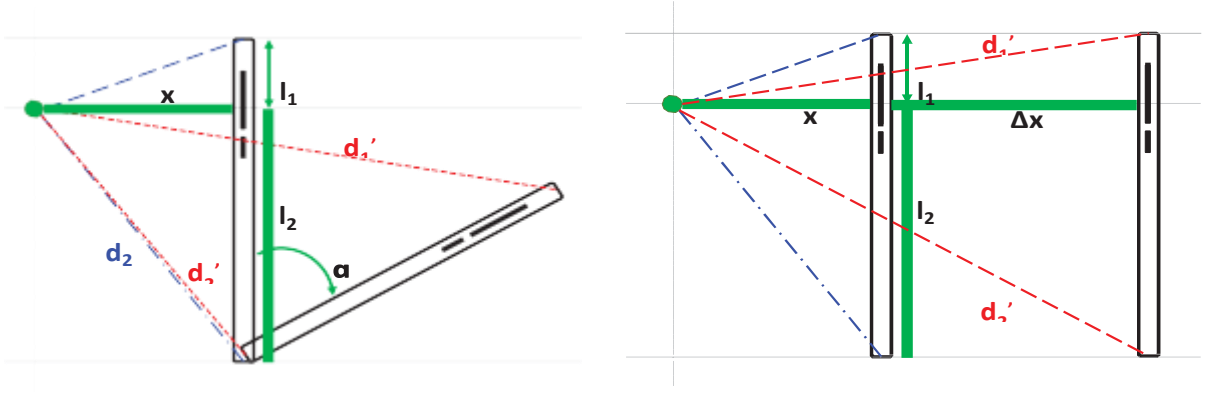

(a) TDoA with angle change (b) TDoA with distance change

**Fig. 8: TDoA changes with different smartphone location and angles.**

We could observe that the English phonemes could be well identified with the distributed TDoA mean values. It is also worth noting that different phonemes' TDoAs process distinct levels of stabilities, which can be related to their articulation positions and manners. Overall, the vowels' TDoAs are more stable than that of the consonants. Especially, voiced consonants like [b], [d] and [g] are all articulated with the stop manner. Such monotonous articulation brings them similar TDoA values and relatively high stability. By contrast, phonemes such as [m], [n] and [U] have lower stability as the articulation positions of these sounds may vary slightly due to different contexts. Moreover, some other phonemes like [k] and [q] yield lowest stability since these voiceless consonants contains no formants but quiet noise-like sounds.

Inspired by such stability distribution, we improve our system performance by increasing the importance of the stable phonemes whereas mitigating the contribution of the less stable ones. Moreover, as the high stability of most phonemes' TDoAs, we achieve text-independent liveness detection by requiring the user to enroll with phonetic sounds instead of specific passphrases.

### 2.5 TDoA Geometric Model with Different Phone Positions

As we could notice, the TDoA measurements change when the position of the smartphone changes. However, it could inconvenient for the user to always hold the smartphone close to his mouth with the same angle and distance. Therefore, we discuss the TDoA changes with different positions of the smartphone.

Specifically, we build geometric models to analyze the TDoA changes with different angles and distances of the smartphone as shown in Figure 8. According to Figure 8(b), we could describe the TDoA after the angle change as:

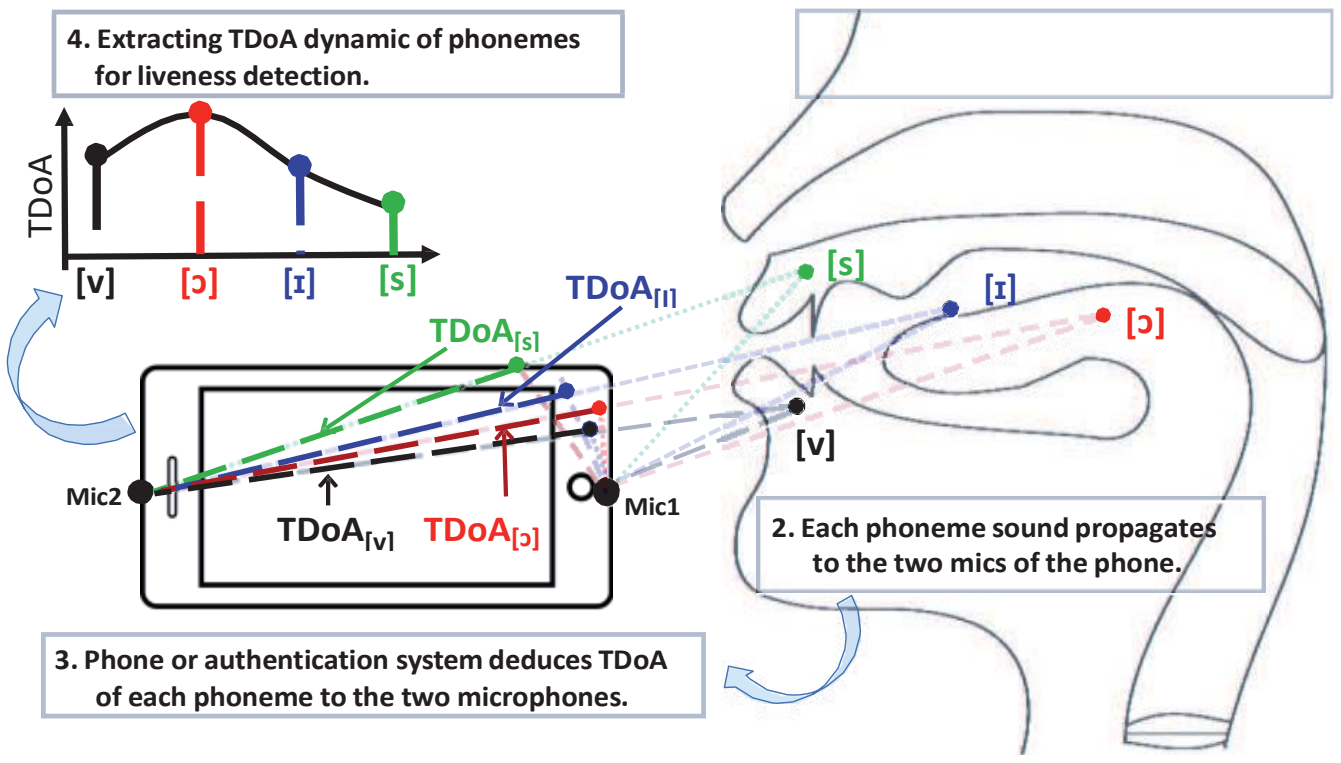

**Fig. 9: Illustration of phoneme localization using a single phone.**

$$\sqrt{l_1^2 + x^2} - \sqrt{l_2^2 + x^2} = TDoA_1 \quad (1)$$

$$\sqrt{l_1^2 + x^2} - \sqrt{(l \cdot sin\alpha + x)^2 + (l_2 - l \cdot cos\alpha)^2} = TDoA_2 \quad (2)$$

where $l$ is the length of the smartphone, $l_1$ and $l_2$ denote the vertical distances from the phoneme origin to the top and the bottom of the smartphone respectively. To simplify this problem, we assume $l_1$ and $l_2$ are known variables, and they are the vertical distances from the user's mouth to the top and bottom of the smartphone. The unknown parameter $x$ denotes the horizontal distance between the phoneme origin and the phone. Parameters $d_1$ and $d_2$ are the distances from the phoneme origin to the top and the bottom of the smartphone respectively. We adopt the gyroscope to measure the angle change $\alpha$, and therefore represent the new TDoA with Equation 2. We solve Equation 1 and 2 to derive two unknown variables, i.e. $x$ and $TDoA_2$, and use the $TDoA_2$ as our new profile.

Similarly, we could describe the TDoA changes with different distances of the smartphone as:

$$\sqrt{l_1^2 + x^2} - \sqrt{l_2^2 + x^2} = TDoA_1 \quad (3)$$

$$\sqrt{l_1^2 + (x + \Delta x)^2} - \sqrt{l_2^2 + (x + \Delta x)^2} = TDoA_2 \quad (4)$$

where $\Delta x$ denotes the distance change between the smartphone and the user. To resolve these equations to get the new TDoA profile $TDoA_2$, we estimate $\Delta x$ by measuring the arrival time of the ultrasound reflection from the user's face [40].

## 3 System Design

In this section, we introduce our system design and its core components and algorithms.

### 3.1 Approach Overview

The key idea underlying our liveness detection system is to perform TDoA ranging for a sequence of phoneme sounds at the two microphones on the phone. As illustrated in Figure 9, a user first speaks an utterance, say “voice" to the phone that closely placed to the user’s mouth. Each phoneme sound (i.e., [v] [A] [I] [s] in the example) is then emitted from the user’s vocal system and picked up by the two microphones of the phone with stereo recording. The phone processes the recorded sound to deduce the TDoA of each phoneme sound to the two microphones. As most phoneme sounds have measurable TDoA differences to the two microphones, a sequence of phonemes will produce series of TDoA with various values, as shown in Figure 9. We refer to the changes in TDoA measurements as “TDoA dynamic", which is then used for liveness detection.

In particular, for the text-dependent system, the measured TDoA dynamic will be compared with the one extracted when the user enrolled in the system. For the text-independent system, the user enrolls in the system with phonetic sounds. During the liveness detection phase, the system searches for the templates of the adopted phonetic sounds in the user’s speech, and then combines these templates as the speech template. Specially, considering the user could hold the device with different positions, we conduct a geometric conversion on the speech template before comparing it with the measured TDoA dynamic. A live user is detected if the similarity score exceeds the predefined threshold. Under playback attacks, the measured TDoA dynamic will be very different from that of a live user due to different sound production systems (i.e., loudspeaker v.s. human vocal system). Under replace attacks, it is extremely unlikely, if not impossible, for an adversary to place a stereo recorder (e.g., smartphone) very close to the victim’s mouth to collect voice samples. Due to the origins of the phoneme sounds are crowded in the mouth and nasal cavities as shown in Figure 5 (b), the TDoA dynamic diminishes rapidly with bluelarge distance between the recorder and the user’s mouth. For example, if the phone is placed 30cm away from the user’s mouth, the maximum achievable TDoA range among all phonemes is less than 1cm. With such a small range, most phonemes have the same TDoA measurement to the two microphones of the phone. The measured TDoAs under replace attack thus cannot match the one extracted when the user enrolled in the system.

Virtually all smartphones are equipped with two microphones and are capable of stereo recording. By leveraging a sequence of phoneme sounds in an utterance/passphrase, our approach relaxes the problem of locating each phoneme sound to tracking TDoA dynamic for live user detection. We thus enable the phoneme localization based liveness detection on a single phone without requiring any additional hardware.

Our system releases the constraints of phone position by conducting geometric conversion on the templates recorded when the user hold the smartphone close to the mouth. The effects of different phones and phone displacement are studied in experiment evaluation. Moreover, data protection mechanisms or secure communication protocols should be in place to prevent an attacker from obtaining the plaintext of TDoA dynamic and the dual-channel audio samples [22]. For example, TDoA dynamic could be extracted locally without storing the dual-channel audio sample, and only the encrypted one-channel audio sample together with the encrypted TDoA dynamic are transmitted or used for verification and liveness detection.

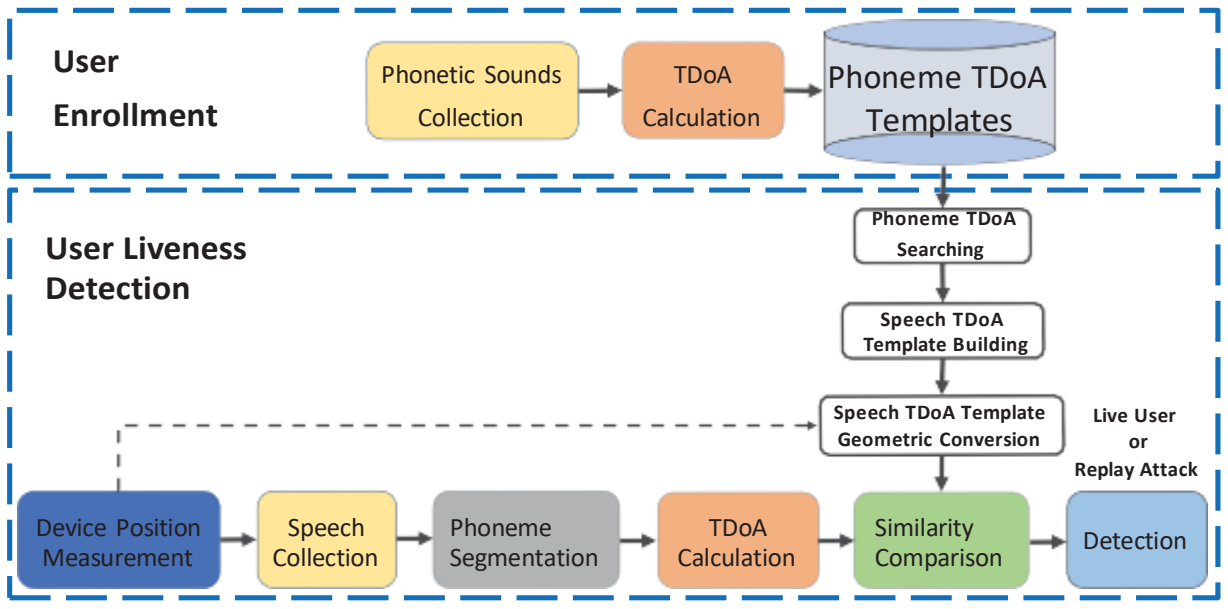

**Fig. 10: The flow of our liveness detection system.**

### 3.2 System Flow

Realizing our system requires five major components: *Device Position Measurement*, *Phoneme Segmentation*, *TDOA Calculation*, *Similarity Comparison*, and *Detection*. As shown in Figure 10, before speech collection, VoiceLive first measures the device position including the phone distance and angle against the user. Nest, the speech acquired by two microphones passes through phoneme segmentation, which extracts phonemes existing in the voice sample. In particular, we combine Hidden Markov Modeling techniques to perform forced alignment on the words recognized from the voice sample to identify each phoneme sound. The words in the voice sample are recognized by acoustic modeling and language modeling algorithms.

Next, the TDOA calculation component is used to calculate the number of delayed samples of each phoneme sound to the two microphones. As acoustic signals can be easily distorted due to multipath propagation, simply correlating phonemes between two channels will result in large error. To address this challenge, we adopt generalized cross-correlation and heuristic-based phase transform weighting approaches for accurate TDoA estimation.

After that, for the text-dependent system, the similarity comparison component measures the similarity of the calculated TDoA dynamic to the one stored in the system. For the text-independent system, VoiceLive builds the speech TDoA template via combining specific the phoneme TDoA templates. Moreover, to release the constraint of different phone positions, we further conduct a geometric conversion on formed speech TDoA template based on the measured device distance and angle. We then compare the TDoA template with the calculated TDoA dynamic. The comparison results in a similarity score, which is then compared with a pre-defined threshold. If the score is larger than the threshold, a live user is detected, otherwise a replay attack is declared. The detection result can be then combined with the traditional voice authentication system to verify the claimed identity of a user.

### 3.3 Device Position Measurement

To measure the device position, we calculate the device distance by timing the ultrasound reflections from the users the face whereas read the device angle via the built-in gyroscope. Specifically, while the user moves the smartphone forward, the built-in loudspeaker of the smartphone keeps sending an inaudible 18-23 kHz beep signal, while the bottom microphone keeps listening and recording the reflected signal. We choose this frequency range as most state-of-the-art smartphones support up to 24 kHz replay and record, as well as to keep the measurement procedures transparent to the user. Moreover, to mitigate the multipath effect whereas ensure the signal receiving and correlation calculation, we choose the signal length as 50 ms, which is 9600 samples with 192 kHz sampling frequency. Besides, considering the user's face is normally within 1 m from the smartphone, we set the interval between each beep signal as 10ms. We calculate the correlation between the original beep signal and the reflected one to decide the distances between the smartphone and the reflectors in the surroundings. Indeed, when the user holds the smartphone, his face is normally always the closest reflector of the beep signal, thus we identify the second peak (the first peak is a result of sounds transmitting through smartphone body) as the reflection caused by the user, and therefore calculate the transmission distance from the device to the user.

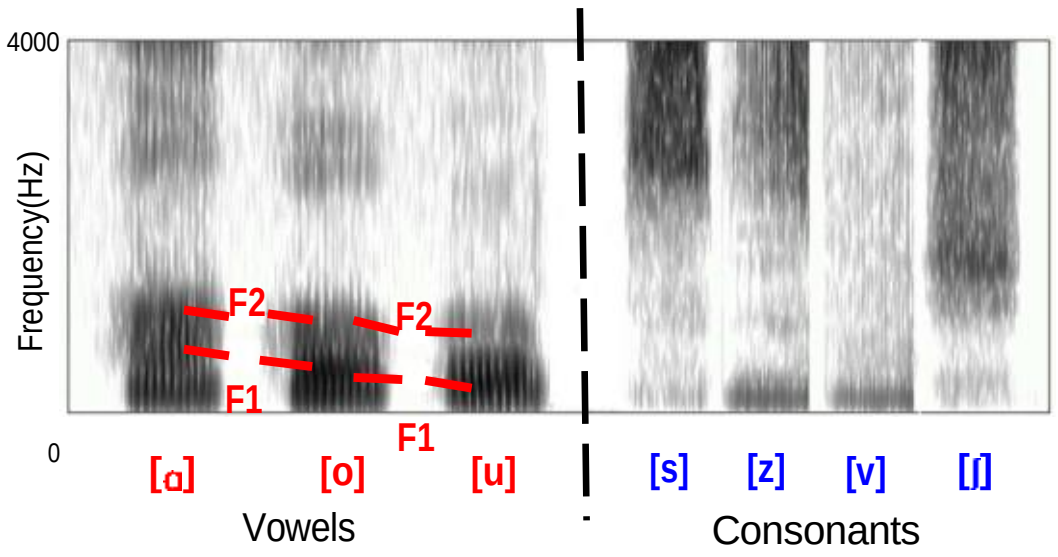

**Fig. 11: Example: spectrogram of phonemes.**

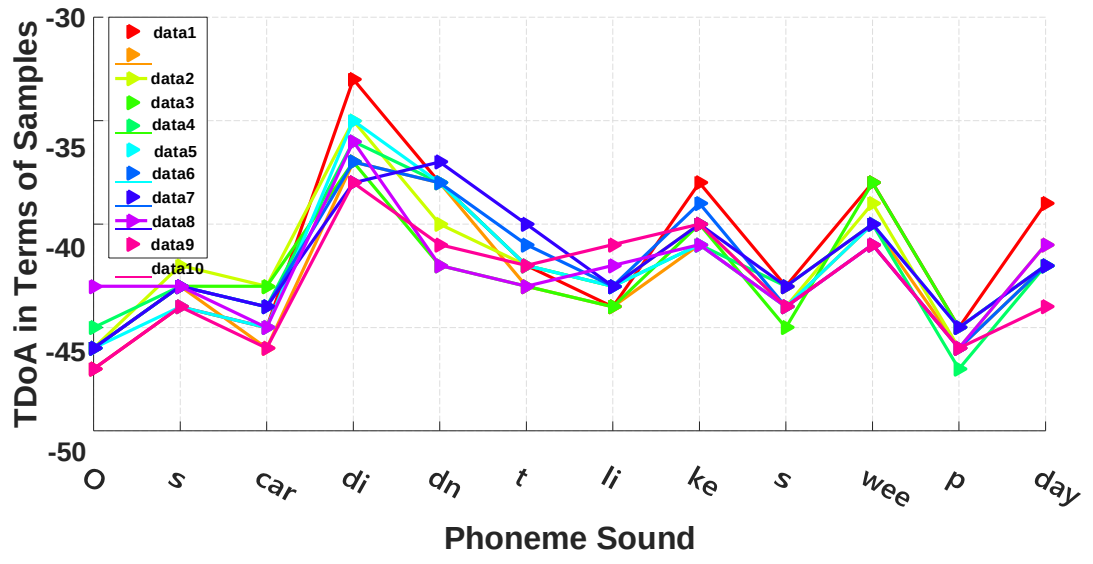

**Fig. 12: TDoAs of one passphrase for 10 trials.**

### 3.4 Phoneme Segmentation

The underlying principle for phoneme segmentation is that the sound of a phoneme contains a number of different overtone pitches simultaneously, known as formants [27]. By analyzing the sound spectrogram, we are able to discover these overtone pitches or formats to identify each individual phoneme sound. Although the most informative formants are the first three formants, the two first formants, F1 and F2, are enough to disambiguate the vowel. As illustrated in Figure 11, it is easy to observe the first two formants, F1 and F2, which contribute to the overtone of each vowel most. It is thus feasible to segment different vowels by looking at the F1 and F2 in the spectrogram. Unlike vowels, consonants' spectrograms display as random mixture of different frequencies, as showed in Figure 11. This static noise-like sound makes it difficult to accurately identify each consonant by simply utilizing formants. We thus adopt forced alignment by using HMM (Hidden Markrov Models), which aligns the input voice spectrogram with existing voice samples to distinguish different consonants [24].

In particular, we first recognize the words existing in the voice sample, which could be done by using automatic speech recognition (ASR). We use advanced CMUSphinx [33] to automatically recognize each word in the user's voice sample. More specifically, the voice sample is first parsed into features, which are a set of mel-frequency cepstrum coefficients (MFCC) that model the human auditory system. Then, the MFCCs are combined together with the dictionary, acoustic model, and language model to recognize the words in the voice sample [33].

Given the recognized words, we utilize MAUS as primary method for phoneme segmentation and labeling [25]. In particular, the recognized words are first transferred into expected pronunciation based on standard pronunciation model (i.e., SAMPA phonetic alphabet). Then, the generated canonical pronunciation together with the millions of possible accents of users yield a probabilistic graph including all possible hypotheses and the corresponding probabilities. At last, the system searches the graph space for the path of phonetic units that have been spoken with highest probability using a Hidden Markrov Model. Outcomes of the search are segmented and labeled phonetic units.

### 3.5 TDOA Calculation

The basic idea of TDoA calculation is to count the number of delayed samples to the two microphones by correlating each segmented phoneme sound between smartphone's two channels. Let's denote $mic_1$ and $mic_2$ as the two microphones/channels of the phone, and $\Delta t$ as the TDoA of one phoneme sound to the two microphones. Given the phoneme sound $mic_1(t)$ recorded at $mic_1$, we correlate such phoneme sound to the sound signal $mic_2(t+d)$ recorded at the $mic_2$, with $d$ varying from 0 to $N-1$. Once the best match is found, the corresponding $d$ value is the number of delayed samples between $mic_1$ and $mic_2$. In particular, such correlation can be done by using a cross-correlation technique [29], as shown below:

$$CC(d) = \frac{\sum_i [(mic_1(i) - \overline{mic_1(i)}) * (mic_2(i+d) - \overline{mic_2(i+d)})]}{\sqrt{\sum_i (mic_1(i) - \overline{mic_1(i)})^2}\sqrt{\sum_i (mic_2(i+d) - \overline{mic_2(i+d)})^2}}, \tag{5}$$

The TDoA $\Delta t$ can be obtained as:

$$\Delta t = \operatorname*{argmax}_d CC(d), \tag{6}$$

However, simply applying the cross-correlation method results in an inaccurate estimation of $\Delta t$ due to the multipath propagation and reverberation effect of acoustic signals. To improve the accuracy, we further utilize generalized cross correlation with phase transformation techniques (PHAT) [26]. By adding a weighting function into cross correlation calculation process, it suppresses the frequency components whose power spectra carry intense additive noises. Meanwhile, PHAT utilizes the cross-power spectral density of two different acoustic signals to improve the system's robustness to reverberation effect. Existing work has shown PHAT can further mitigate the spreading effect that caused by uncorrelated noises at two microphones [26].

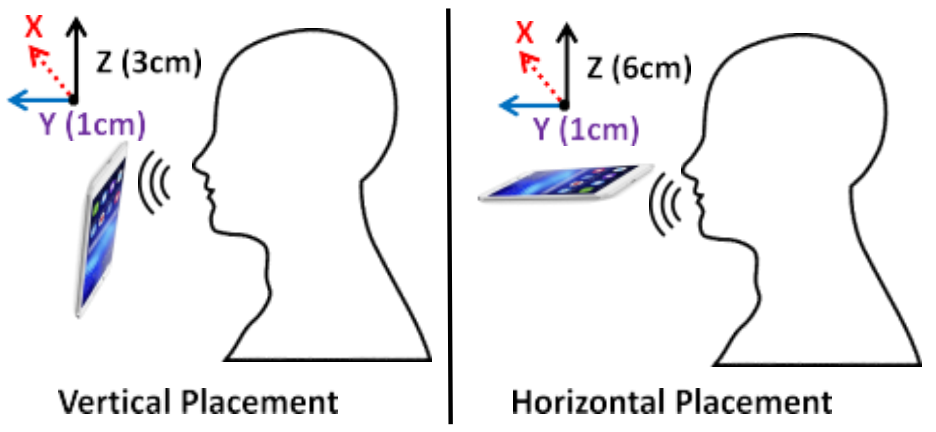

**Fig. 13: Two different phone placements diagram.**

Figure 12 shows one example of the TDoA values when one participant performs 10 trials of authentication with the passphrase "Oscar didn't like sweep day". The X axis shows each phoneme sound, whereas Y axis shows the TDoA in terms of number of delayed samples. We observe that TDoA dynamics of these trials are highly similar and stable, with only 1 to 2 samples variation under 192kHz sampling rate. The results show that TDoA calculation is able to catch the user's unique speech production system accurately.

### 3.6 Similarity Comparison of Text-dependent System

Once the TDoA dynamic is extracted, we first normalize these TDoA values to the same scale as those stored in the user profile. Such normalization is used to deal with the issues of device diversity and phone displacement. The phone a user used to enroll in the system could be different from the one he/she used for authentication. As different phones differ in size or distance between the two microphones, the absolute TDoA values of the same phoneme could be different. Similarly, if the user places the phone at a location slightly different from that when he/she enrolled in the system, the absolute TDoA values vary slightly. Normalizing the TDoAs to the same scale could effectively mitigate these issues.

To compare the similarity of the TDoA dynamic with the user profile, we utilize both the correlation coefficient and the probability. In particular, the correlation coefficient measures the degree of linear relationship between two sequences [43]. Other than calculating the absolute difference, it quantifies the similarities in the changes of two sequences. The correlation coefficient ranges from -1 to +1. A value close to +1 indicates a high degree of similarity, whereas a value near 0 indicates a lack of similarity [47].

For the probability based method, we assume the TDoA ranging error of each phoneme follows an independent standard Gaussian distribution. Given the TDoA value $TDoA_i$ in the extracted TDoA dynamic, the probability that it matches the one in the user profile is represented as:

$$P(TDoA_i) = \frac{1}{\sigma\sqrt{2\pi}} e^{-(TDoA_i - \overline{TDoA_i})^2} \quad (7)$$

whereas $\sigma$ is the standard deviation of the error and $\overline{TDoA_i}$ is the corresponding TDoA value in the user profile. During the user enrollment phase, we ask each user to speak a passphrase three times to extract the averaged TDoA and the standard deviation of each phoneme for similarity comparison. Given the probability value of each phoneme, we simply average the probability values of all phonemes as the indicator of the similarity score.

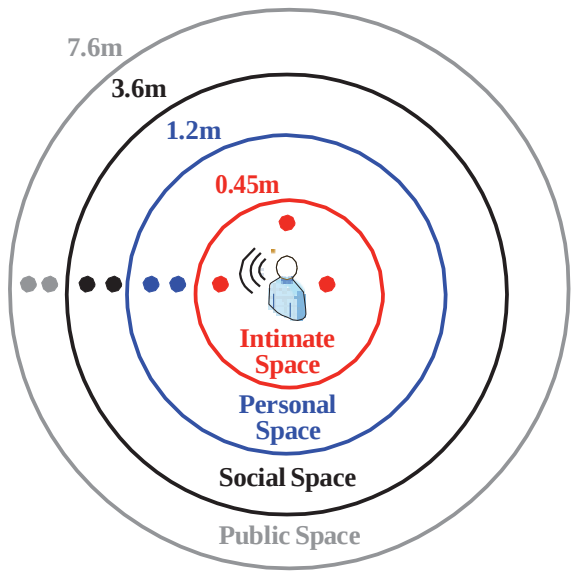

**Fig. 14: Illustration of locations of replace attacks and different types of social distances.**

Correlation coefficient and probability are two metrics targeting on different characteristics of the TDoA dynamic. We refer to the former as *Correlation*, and latter as *Probability*. Moreover, we develop a combined scheme that simply combines the similarity scores of the correlation and probability based methods. We refer to such a method as *Combined method*, which takes advantages of both the correlation coefficient and the probability.

### 3.7 Similarity Comparison of Text-independent system

We further extend our liveness detection system to be text-independent. Instead of comparing the TDoA dynamic of the input passphrase with that of the enrolled one directly, we build general TDoA profile models for each user, based on his pronunciations of all the 44 English phonemes. Before similarity comparison, we search for the TDoA profile of each phoneme in the passphrase, and form the TDoA dynamic profile of the given passphrase by combining all involved phonemes' TDoA profiles. Indeed, a text-dependent liveness detection system requires the user to enroll in any passphrases that used for liveness detection and it is only valid on these enrolled speech. In comparison, a text-independent liveness detection system functions independently of passphrases' contents.

This is feasible, since according to our preliminary study on the TDoA stability distribution of English phonemes, most English phonetic sounds' TDoAs are stable enough regardless of the different contexts. Moreover, we utilize the different stabilities of the TDoA values as weights to further improve the detection accuracy of our text-independent system by the following equation:

$$\gamma_{xy} = \frac{\sum_{i=1}^{n}[w_i(x_i - \bar{x})(y_i - \bar{y})]}{\sqrt{\sum_{i=1}^{n}[(w_i(x_i - \bar{x}))^2]}\sqrt{\sum_{i=1}^{n}[(w_i(y_i - \bar{y}))^2]}} \quad (8)$$

For a specific phoneme $x_i$ in a sequence of n phonemes, we find the phoneme profile of $x_i$ and set it as $\bar{x}$, and then we adopt the group std as the $w_i$. We still employ (7) for probability measurement. However, for the text-independent system, the $\sigma$ is the standard deviation of the given phoneme and $\overline{TDoA_i}$ is the corresponding TDoA profile of that phoneme.

This method allows to compare the current TDoA dynamic with a phoneme based TDoA profile, rather than a specific passphrase based TDoA profile. Therefore, it enable our system to conduct liveness detection continuously during the whole user-device communication session, regardless of text or passphrase contents.

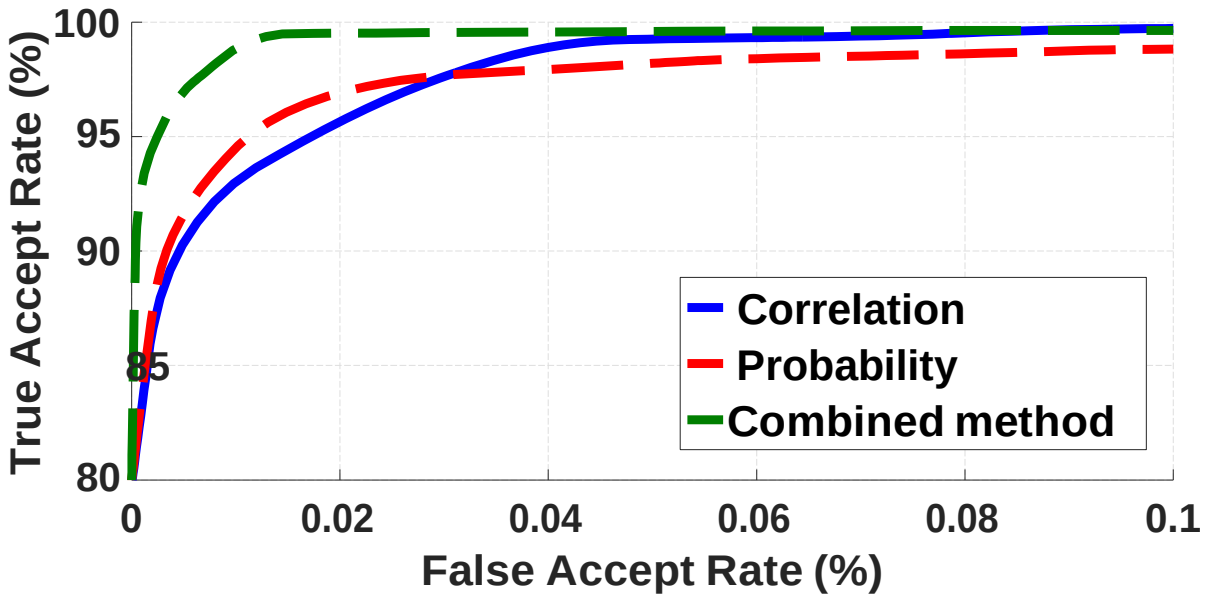

**Fig. 15: Playback Attacks: ROC curves under different methods.**

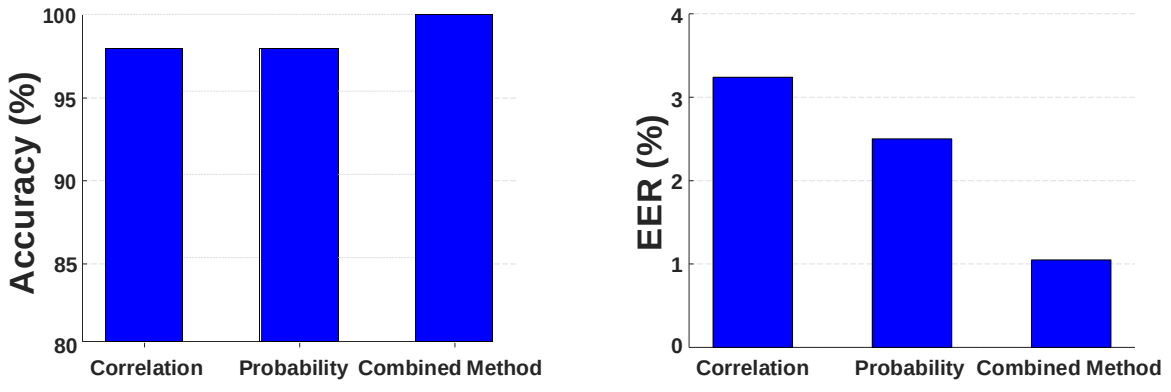

(a) Accuracy (b) EER

**Fig. 16: Playback Attacks: Accuracy and EER.**

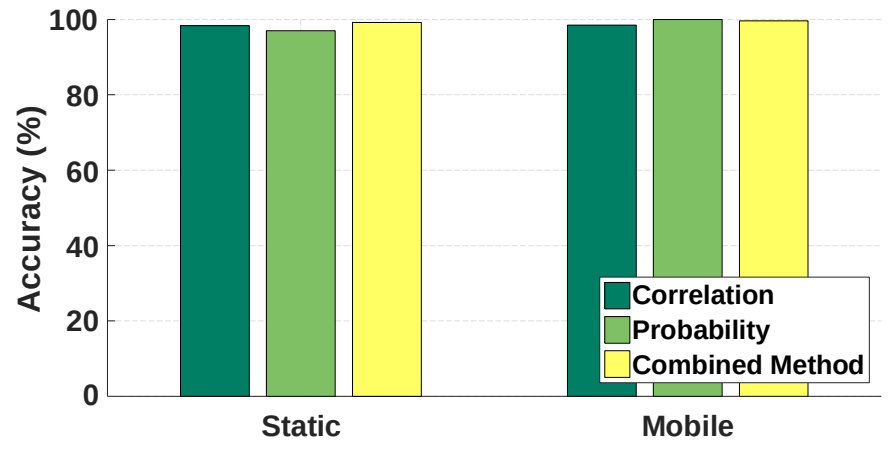

**Fig. 17: Static and Mobile Playback Attacks: Accuracy under different methods.**

## 4 Performance Evaluation

In this section, we evaluate our liveness detection system under replay attacks including both *playback* and *replace* attacks[3]. We also evaluate the robustness of our system to different types of phones, sampling frequencies, phone displacements, and lengths of passphrases.

### 4.1 Experiment Methodology

**Phones and Placements.** We evaluate our system with three types of phones with different sizes and audio chipsets. In particular, we experiment with Samsung Galaxy Note3, Galaxy Note5 and Galaxy S5. The distance between the two microphones (i.e., one on the top and one at the bottom) for stereo recording is about 15.1cm for Note3, 15.3cm for Note5, and 14.1cm for S5. The audio chipset of Note3 is Qualcomm Snapdragon 800 MSM8974, whereas it is Wolfson WM1840 for Note5, and Audience's ADNC ES704 for S5. The operating system of these phones is Android 6.0 Marshmallow, which enables the phones to perform stereo recording at 48kHz, 96kHz and 192kHz sampling frequencies. These frequencies represent ranging resolutions of 7.08mm, 3.54mm, and 1.77mm, respectively. We use 192kHz as our primary sampling frequency and present the corresponding results unless otherwise stated. We also experiment with two types of phone placements, as shown in Figure 13. One is vertical placement with the phone placed close to user's mouth vertically. We call such placement our primary placement and present the performance of such placement unless otherwise specified. For the vertical placement, the phone is about 3cm and 1cm away from user's mouth on Z and Y axis, respectively. The other one is horizontal placement with the phone placed close to the user's mouth horizontally. The phone is about 6cm and 1cm away from user's mouth on Z and Y axis, respectively.

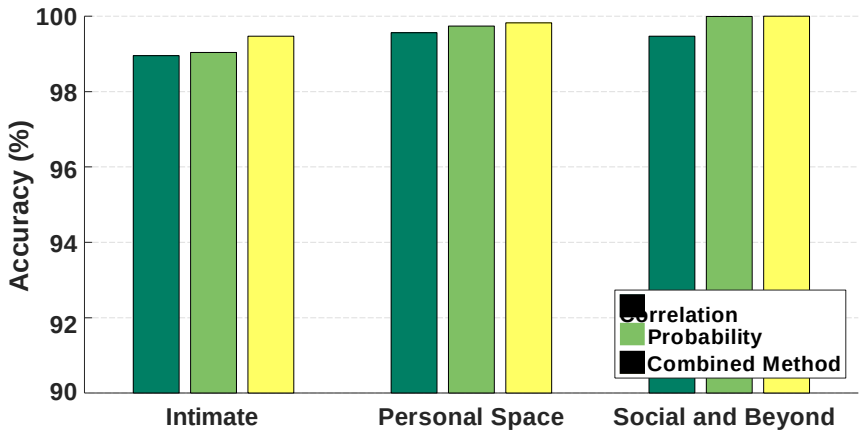

**Fig. 18: Replace Attacks: Accuracy of different methods with different social distances.**

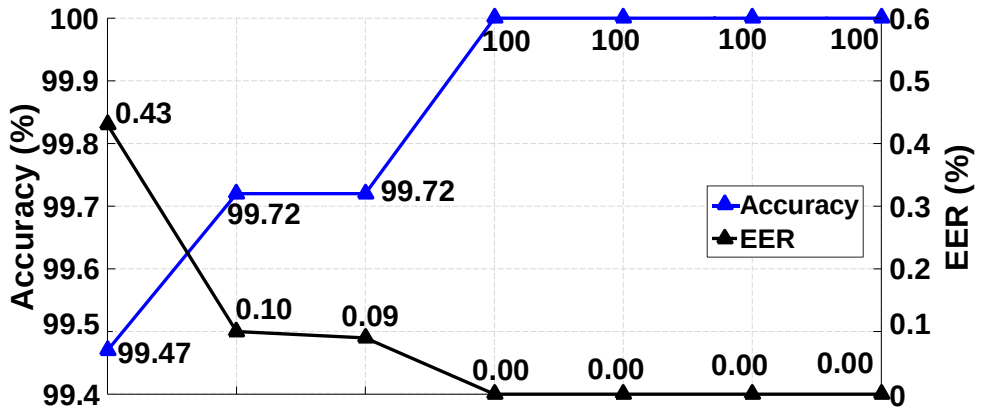

**Fig. 19: Replace Attacks: EER and Accuracy of Combined method under different distances.**

**Data Collection.** Our experiments involve 12 participants including 6 males and 6 females whose ages range between 25 to 38. These participants are either graduate students or university researchers, who are recruited by emails. The participants are informed of the purpose of our experiments and are required to act as if they were conducting voice authentication. Each participant chooses 10 different passphrases of their own and performs 10 times legitimate authentications for each passphrase after enrollment. To enroll in the system, each participant speaks a passphrase three times to extract the averaged TDoA and the standard deviation of each phoneme for similarity comparison. For online verification, users only speak the passphrase once. Each participant speaks the passphrase with her/his habitual way of speaking. The lengths of the passphrases are ranging from 2 words to 10 words with proximately half of them are 2-4 words, one quarter of them are 5-7 or 8-10 words. The experiments are conducted in both the office and home environments with background and ambient noises, such as people chatting and HVAC noise.

**Attacks.** We experiment with two types of replay attacks: *playback attacks* and *replace attacks*. For playback attacks, we replay participants' voice samples in front of the smartphone that performs stereo recording for authentication. We utilize three different types of loudspeaker including DELL AC411 wireless speaker system, Samsung Galaxy note5 and S5 speakers, to replay each pre-recorded voice sample. In addition, half of the playback attacks are conducted with stationary loudspeakers that are within 10cm away from the smartphone (i.e., *Static Playback Attacks*); while the other half

3. This project has obtained IRB approval.

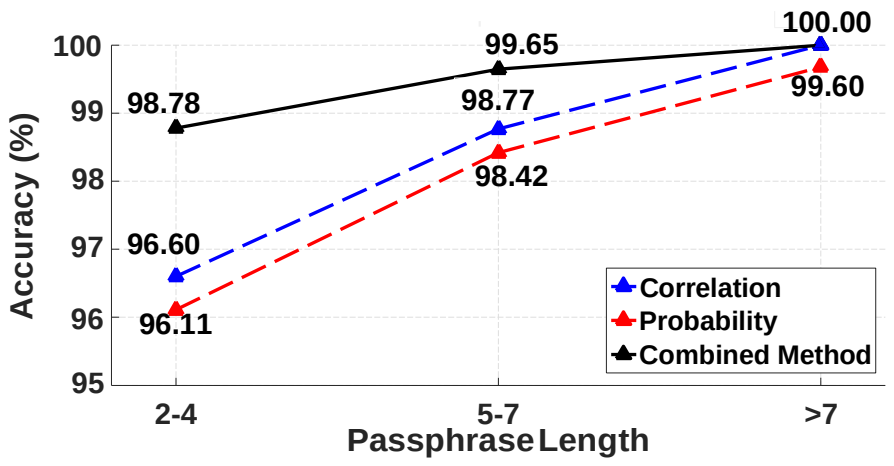

Fig. 20: Accuracy under different lengths of passphrase.

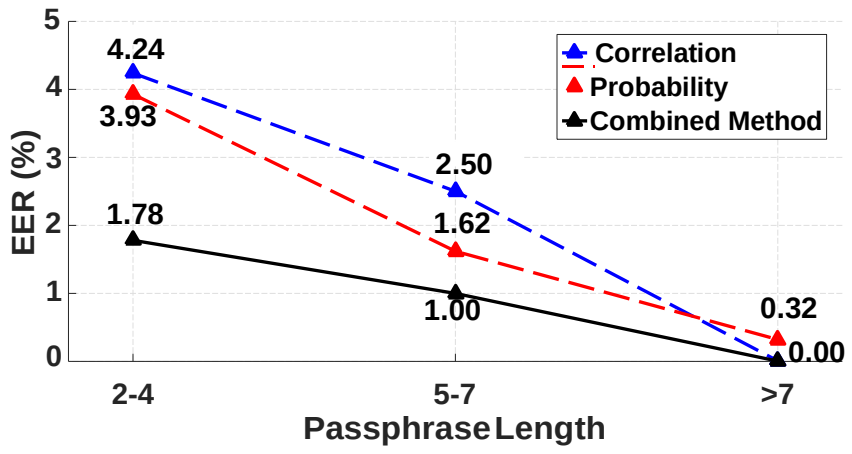

Fig. 21: EER under different lengths of passphrase.

are conducted with mobile loudspeakers targeting on mimicking TDoA changes of users by moving the loudspeakers around the smartphone (i.e., *Mobile Playback Attacks*).

For replace attacks, we place a smartphone with stereo recording close to the target user when the user is performing legitimate voice authentication. In such cases, the adversary obtains a two-channel voice sample of the target and then uploads that directly to the voice authentication system. The only difference between the two-channel voice sample obtained by the adversary and the one in legitimate authentication is the recording distance. We adopt the Edward T. Hall's proxemics theory [19] to emulate how close an adversary could place the phone next to the user's mouth. As shown in Figure 14, the minimum distances between people are categorized by the relationship and types of interactions between them. It includes intimate distance, personal distance, social distance, and public distance. With such a guideline, we chose the recording distances between the attacker's phone to the user's mouth as 30cm, 50cm, 100cm, 150cm, 200cm, 300cm, and 450cm, which simulates different types of relationships. We also consider the circumstances where the attacker could hide behind or at the side of the user. The recording distances for such cases are limited by the size of user's head, and are around 40 cm and 25 cm away to user's mouth, as shown in Figure 14.

**Metrics.** We use the following metrics to evaluate the performance of our liveness detection system. *False Accept Rate (FAR)*: the probability that the liveness detection system incorrectly declares a replay attack as a live user. *False Reject Rate (FRR)*: the probability that our system mistakenly classifies a live user as a replay attack. *Receiver Operating Characteristic (ROC)*: it describes the relationship between the True Accept Rate (i.e., the probability to identify a live user as a live user) and the FAR when varying the detection threshold. *Equal Error Rate (EER)*: it shows a balanced view of the FAR and FRR and is defined as the rate at which the FAR equals to the FRR. *Accuracy*: it measures the overall probability that the system could detect a live user and reject a replay attack.

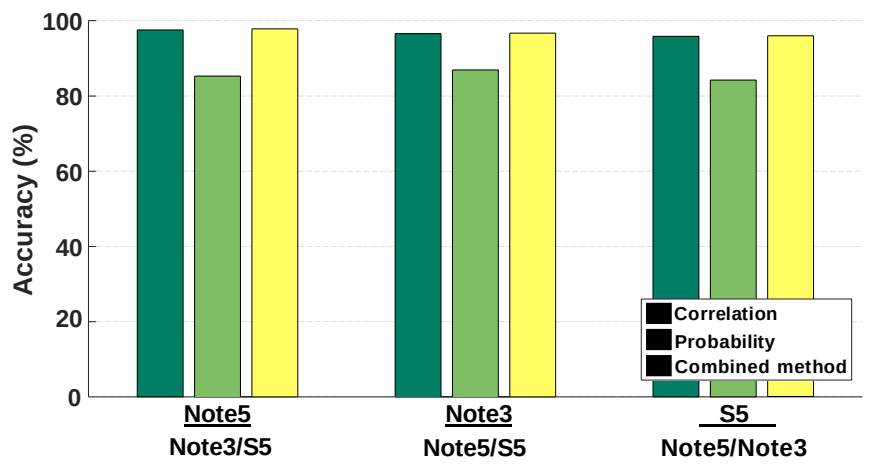

Fig. 22: Accuracy of using one phone as enrollment and the other two as online authentication.

### 4.2 Overall Performance of Text-dependent System

We first evaluate the overall performance of our liveness detection system under two types of replay attacks: *playback attacks* and *replace attacks*.

**Playback Attack.** Figure 15 shows the ROC curves of different methods in detecting live users under playback attacks. We observe that our system is highly effective in detecting live users and rejecting playback attacks. These three methods provide more than 94% detection rate with less than 1% FAR. In particular, the correlation and probability based methods have comparable performance. The correlation method provides a detection rate of 95% with 1% FAR. The combined method has the best performance and results in over 99% detection rate with less than 1% false accept rate.

Moreover, Figures 16 depicts the overall accuracy and EER of different methods under playback attacks. We observe that the combined method provides the best accuracy and EER, which are 99.30% and 1.05% respectively. The correlation method produces an accuracy of 97.95%, which is slightly better than that of the probability method (i.e., 97.54%). However, probability method results in a better EER than that of the correlation method. In particular, probability method has an EER of 2.50% and correlation method has an EER of 3.24%. The above results show that VoiceLive is highly accurate in detecting live users under playback attacks, and the combined method provides the best results since it takes advantages of both the correction and the probability based methods.

We next take a closer look at how our system performs under static and mobile playback attacks. In our experiments, we observe the static playback attacks produce similar TDoA values for different types of phoneme sounds. Although playback attacks under mobile scenarios could result in TDoA changes, the resulted changes in TDoA cannot match with the ones in the user profile. It is because the attacker couldn't mimic the sound position transition the same as that of the human vocal system. As shown in Figure 17, our system is highly effective in live user detection under both static and mobile playback attacks. The combined method achieves 99.2% accuracy under static scenarios and 99.65% accuracy under mobile cases.

**Replace Attack.** We next evaluate the effectiveness of our system in defending against the replace attacks. Figure 18 illustrates the accuracy of different methods with replace attacks conducted under different social distances. In particular, the testing positions of replace attacks fall into three categories: *intimate* (<45cm), *personal space* (45cm to 1.2m), and *social and beyond*(>1.2m). We observe that our system can effectively detect the live users and reject the replace attacks under each category of social distances. For example, the combined method provides 99.47% detection accuracy under intimate relationship, 99.82% under personal relationship, and 100% under social relationship and public space. And all the methods provide over 98.95% detection accuracy across different categories.

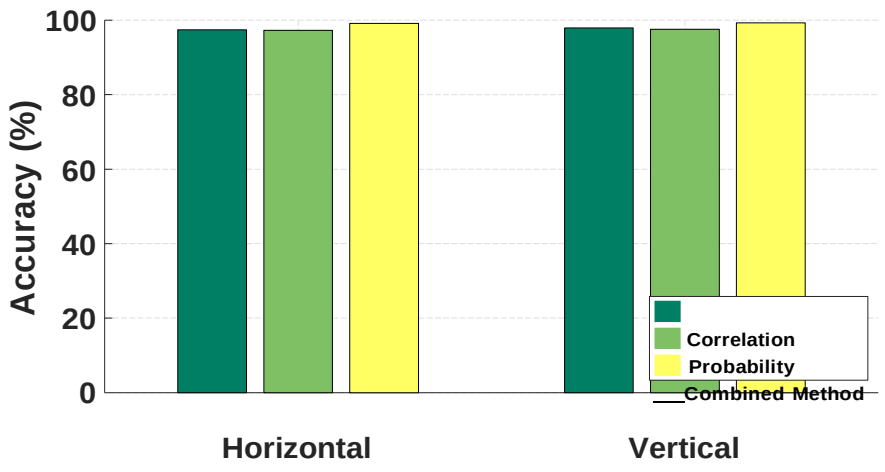

**Fig. 23: Accuracy of horizontal and vertical placements.**

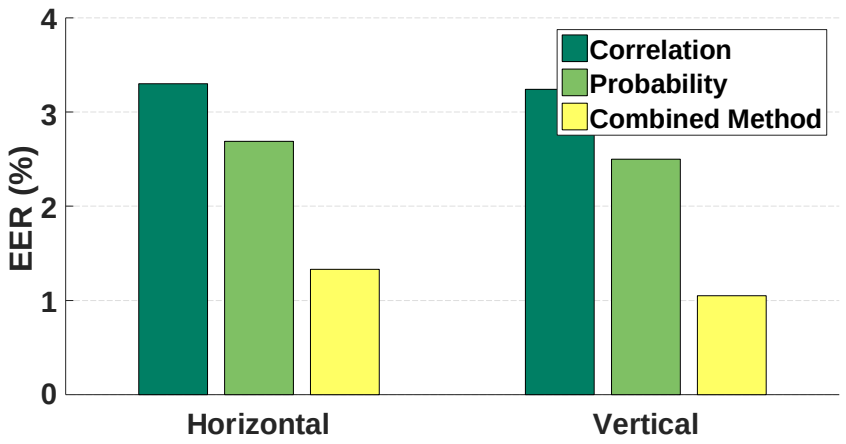

**Fig. 24: EER of horizontal and vertical placements.**

Figure 19 shows the details on the accuracy and EER of the combined method under each social distance. We find that both the accuracy and EER are improved with an increased social distance. In particular, the EER decreases from 0.43% to 0% and the accuracy is improved from 99.47% to 100% when the distance is increased from 30cm to 150cm. When the attacker is further away, our system can detect all the live user cases and reject all the replace attacks. This is because when increasing the distance between the phone and user's mouth, the TDoA dynamic diminishes rapidly.

We also investigate the replace attacks launched from 25cm behind the user and 40cm from the side of user. The EER of the combined method under these two cases are 0.33% and 0%, respectively. Such results are comparable to the EER in Figure 19. This shows our system is capable of detecting replace attack conducted from different directions.

### 4.3 Impact of Passphrase Length

Generally, a passphrase with longer length provides stronger security. It also produces more phoneme sounds that generate more changes in the TDoA measurements. We thus study the performance of our system with different lengths of passphrases. In particular, we sort all the passphrases into three categories including short passphrases with 2 to 4 words, appropriate passphrasses with 5 to 7 words, and long passphrases with 8 to 10 words. Note that researchers and professionals in voice authentication suggest that a passphrase should contain at least 5 words so as to provide sufficient security level [8].

Figure 20 and Figure 21 illustrate the accuracy and EER for different lengths of passphrases, respectively. We observe that both the accuracy and EER are improved for all the methods when we increase the length of the passphrase. In particular, the accuracy is improved from 98.78% to 100% and the EER is reduced from 1.78% to 0% for combined method, when we increase the length from 2-4 words to more than 7 words. In addition, with an appropriate length of passphrase (i.e., 5 to 7 words), the combined method results in 99.65% accuracy and 1% EER. The results confirm our observation that a longer passphrase leads to more TDoA changes of phoneme sounds, which improves the performance of the live user detection.

### 4.4 Impact of Different Phones

As one user may use one phone to enroll in the system but uses another one to perform online authentication, we study how our system behaves under different phones. Specifically, we experiment with users to use either Note5, Note3, or S5 to enroll in the system and then utilize the other two for online authentication. These three types of phones differ in size and audio hardware as described in the experimental setup. Figure 22 shows the accuracy of different methods when using one phone as enrollment and the others as online authentication. We observe that our system still provides accurate liveness detection. In particular, the combined method results in accuracy of 97.82%, 96.67%, 96% when using Note5, Note3, and S5 as the phone for enrollment while the other two for authentication, respectively. Moreover, we observe that these three methods have comparable performance no matter which phone is used for enrollment or authentication. Although the accuracy is slightly higher when using the same phone (i.e., about 99%), our system still produces very accurate detection results with different phones. Such observations show that our system is robust and compatible to different phone models.

### 4.5 Effect of Phone's Placement

Different users might have different ways to place the phone close to their mouths during the authentication process. We thus compare the performance of our system under two types of placements, vertical and horizontal, as described in experimental setup. Figure 23 illustrates the accuracy comparison of these two placements. We observe that our system achieves very high accuracy for both placements, with the accuracy slightly higher when the phone is placed vertically. Specifically, the accuracy under horizontal placement is 97.41%, 97.26%, and 99.13% for correlation, probability, and combined method respectively. Figure 24 shows the EER under two displacements. We have similar observation to that of the accuracy. In particular, EER is 3.3%, 2.69%, and 1.33% for correlation, probability, and combined method respectively. Results show that our system works very well for different phone placements including both horizontal and vertical placements.

### 4.6 Overall Performance of Text-independent System

We compare the overall performance of both text-dependent and text-independent VoiceLive under playback attacks and replace attacks respectively. Figure 25 and Figure 26 present the overall accuracy and EER of the proposed methods under playback attacks. With different methods, the text-dependent system provides 97.75%, 97.54% and 99.30% accuracy whereas 3.24%, 2.50% and 1.05% EER.

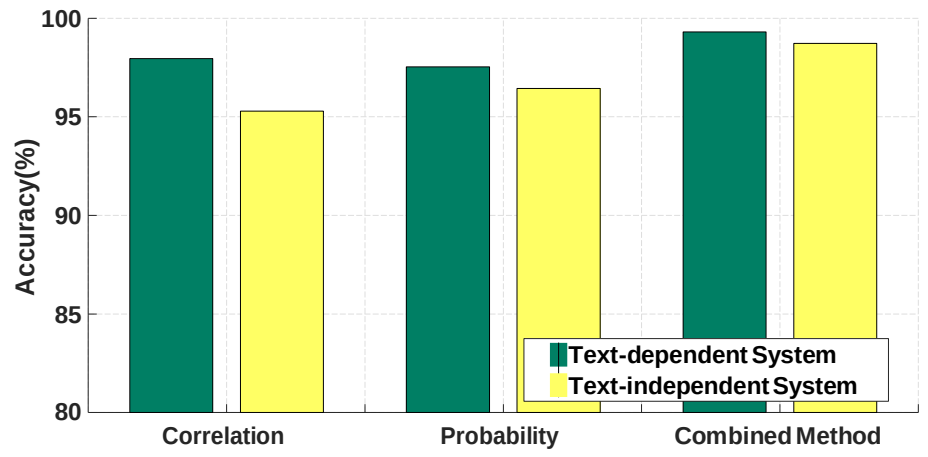

**Fig. 25: Playback Attacks: Accuracy of VoiceLive and VoiceLive+.**

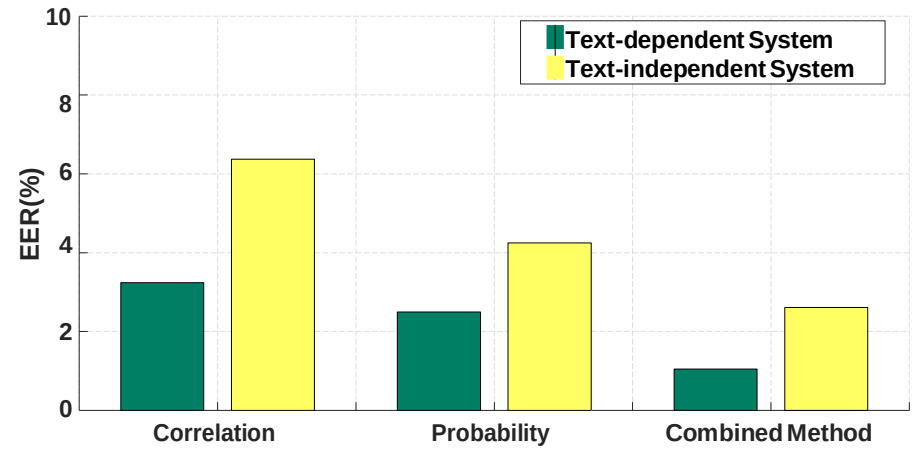

**Fig. 26: Playback Attacks: EER of VoiceLive and VoiceLive+.**

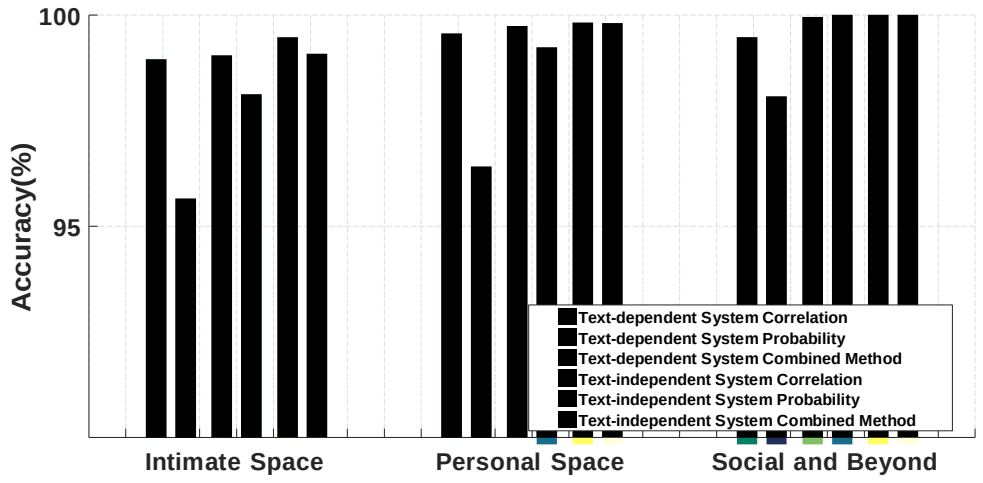

**Fig. 27: Replace Attacks: Accuracy of VoiceLive and VoiceLive+.**

In comparison, the text-independent system yields 95.29%, 96.44% and 98.72% accuracy while 6.37%, 4.25% and 2.61% EER. We notice that, in exchange of releasing the constraint of text-dependence, our system suffers from some performance degradation with the correlation method, nevertheless, both probability and the combined method provide comparable performance with the text-dependent version.

Similarly, Figure 27 illustrates the accuracy of the text-dependent and text-independent systems under different social distances, with different methods. Specifically, for three levels of social distances, the combined method of text-dependent system provides 99.47%, 99.82% and 100% accuracy, whereas the text-independent system yields 99.08%, 99.81% and 100% accuracy. The correlation method is more influenced than the probability method, nevertheless, VoiceLive+ enables to achieve equivalent performance with combined method.

In summary, with the combined method of correlation and probability, the text-independent VoiceLive is able provide proper accuracy in defending both playback and replace attacks, and identifying live users for text-dependent speaker verification systems.

### 4.7 Impact of Releasing Phone Position Constraints

We setup experiments as shown in Figure 28 to explore our system performance with different angles and distances of the smartphone. Specifically, we moves the smartphone forward from the user's mouth for 5 cm, 10 cm, and 15 cm.

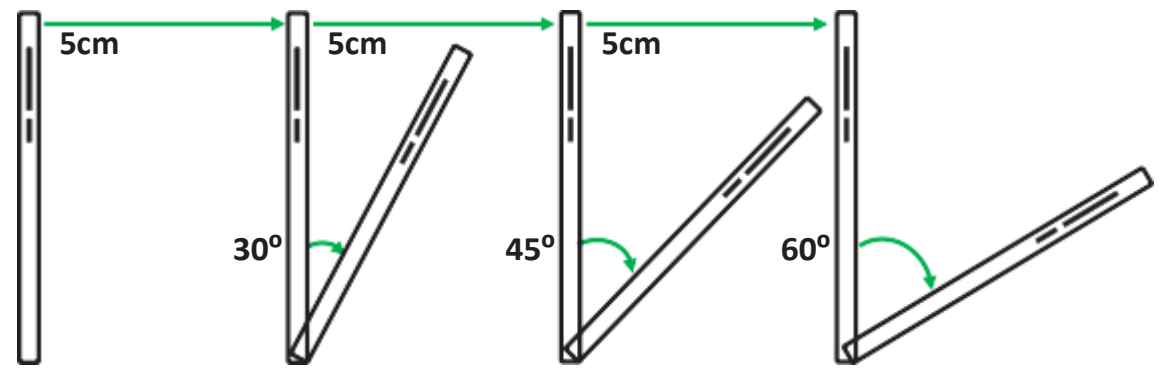

**Fig. 28: Experiment setup.**

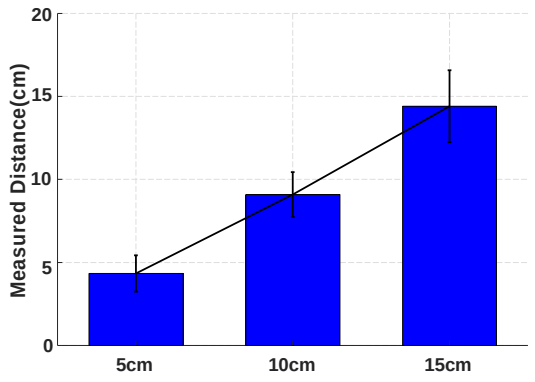

**Fig. 29: Distance measurement error.**

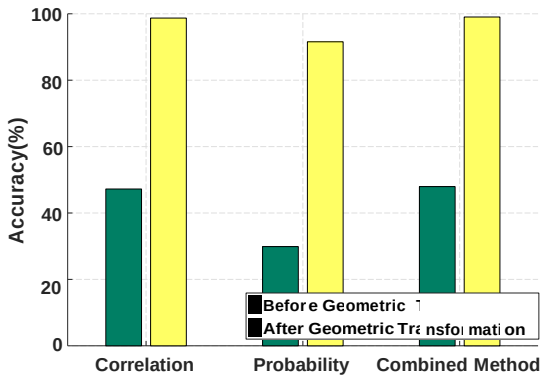

**Fig. 30: Overall accuracy with distance and angle change.**

At each distance, we rotate the smartphone for 30°, 45°, and 60°. Figure 29 shows that with the discussed distance measurement, the measured mean values are 4.34, 9.09, and 14.41 samples; whereas the std values are 1.09, 1.35 and 2.17 samples. Indeed, when the smartphone is further away, the mean value is closer to the ground truth while the std is larger.

Figure 30 displays the accuracy comparison of our system before and after the geometric transformation. As we could observe, without geometric transformation, the system performance jumps to as low as around 50%, whereas after the geometric transformation, the distance and angle changes could not degrade our the system accuracy, which is 98.72%, 91.58% and 99.05% with the correlation, probability, and the combined method respectively. This accuracy is comparable to our system overall performance.

## 5 Related Work

Recent years more and more mobile devices and apps are embracing voice biometric for mobile authentication. For example, Android OS has introduced voice authentication to allow users to unlock their devices [2], [7], whereas Apps like Wechat enable voice logins [9]. Moreover, financial institutions are also integrating voice biometric in their telephone and online banking systems, such as SayPay, HSBC, Citi, and Barclays [3], [4]. However, voice authentication is subject to spoofing attack, as indicated in recent studies [17], [20], [39]. Voice spoofing attacks can be divided into four categories, which are described below together with corresponding countermeasures.

**Replay Attack.** An adversary can effectively spoof a voice authentication system via replay attacks [28]. To defend such attacks, some researchers propose to detect the artifacts introduced by the replay device or the environments. For example, Villalba *et al.* utilize the increased noise and reverberation of replaying far-field recordings for attack detection [37]. Wang *et al.* use the additional channel noise of the recording and loudspeaker for attack detection [39]. Chen *et al.* detect the magnetic fields of the loudspeaker to confirm a replay attack [14]. However, the effectiveness of these approaches is very limited in practice (e.g., the FAR rate could be as high as 17%.) Whereas other work focus on liveness detection, which explores unique characteristics of a speaking human to distinguish live users from replay devices. For instance, Zhang *et al.* monitor specific ultrasound reflections caused by articulators' motions when the user speaks a given passphrase [44] whereas Meng *et al.* adopt WiFi devices to sense such features [30]. Feng *et al.* examines the body vibrations with contact devices like earbuds and eyeglasses [18] while Shang *et al.* try to record such vibrations by attaching a smartphone close to the user's throat [35]. Wang *et al.* measure the exhalation noises in registered passphrases [38]. However, all these solutions are restricted to the distance as they require either additional attached devices, or the user to be extremely close to the device. Moreover, these methods are text-dependent, which only function on enrolled passphrases. As an alternative, Yan *et al.* catch the distinctive sound fields of enrolled users. Despite of text-independent, this work requires the user to hold the smartphone at a fixed location to collect similar field prints [42].

**Impersonation Attack.** It refers to attacks that an adversary tries to mimic the victim's voice without utilizing any computer or professional devices. Recent work shows that impersonation attack could be defended very efficiently by using advanced speaker model, such as GMM-UBM [12] and i-vector models [20]. Existing voice authentication systems with such advanced speaker models thus are resistant to impersonation attacks.

**Speech Synthesize Attack.** This type of attack indicates an attacker has the ability to synthesize victim's voice by utilizing speech synthesize technologies. Earlier work done by Lindberg and Blomberg [28] shows that the FAR can be increased to as high as 38.9% with less sophisticated speaker models. Recent work done by De Leon et al. shows that by adopting both GMM-UBM and SVM technologies, voice authentication systems are able to lower the FAR of the system to 2.5% [17]. Also, Chen et al. [13] show that by employing higher order Mel-cepstral coefficients, the EER can be lowered to 1.58%.

**Voice Conversion Attack.** This kind of attack aims to manipulate or convert existing voice samples from other users so that it would resemble target's voice. In the early work, researchers demonstrate such attacks can significantly affect the voice authentication system [23]. Recently, Wu et al. [41] developed an authentication system with PLDA component that could effectively defend against voice conversion attacks with 1.71% FAR, whereas Alegre et al. utilize PLDA and FA technologies which result in the FAR rate of 1.6% [10].

## 6 Conclusion

In this work, we developed a liveness detection system for voice authentication that requires only stereo recording on smartphones. Our system VoiceLive is practical as no additional hardware is required during the authentication process. VoiceLive performs liveness detection by measuring TDoA changes of a sequence of phoneme sounds to the two microphones of the smartphone. It distinguishes a live user from a replay attack by comparing the TDoA changes of the input utterance to the one stored in the system. Our experimental evaluation demonstrates the viability of distinguishing between a live user and a replay attack under various experimental settings. Also,the distinct traits of vocal systems may be adopted as new biometrics by future authentication systems. Our experimental results also show the generality of our system, as we experiment with different phone types, placements and sampling rates. Overall, VoiceLive can achieve over 99% accuracy, with the EER as low as 1%.